\documentclass[12pt]{spieman}  
\usepackage{amsmath,amsfonts,amssymb}
\usepackage{graphicx}
\usepackage{setspace}
\usepackage{tocloft}
\usepackage{mdframed}
\usepackage{xfrac}
\usepackage{lineno}
\usepackage[most]{tcolorbox}
\usepackage{xcolor}

\definecolor{lightgraybox}{RGB}{245,245,245}

\tcbuselibrary{breakable}

\newtcolorbox{prefixbox}{
  breakable,
  enhanced,
  colback=lightgraybox,
  colframe=gray!50,
  boxrule=0.5pt,
  arc=4pt,
  left=8pt,
  right=8pt,
  top=6pt,
  bottom=6pt,

}

\title{Exact Expressions of Entropy for Classical Non-interacting Many-body Systems}

\author[a,c,*]{Yuheng Wu}
\author[a]{Henrik Heelweg}
\author[b]{Rigel Galgana}
\affil[a]{Department of Chemistry, Massachusetts Institute of Technology, Cambridge, United States}
\affil[b]{Operations Research Center, Massachusetts Institute of Technology, Cambridge, United States}
\affil[c]{Department of Material Science and Engineering, Massachusetts Institute of Technology, Cambridge, United States}

\cftpagenumbersoff{figure}
\cftpagenumbersoff{table} 
\begin{document} 
\maketitle

\begin{abstract}
In the thermodynamic limit, the equilibrium state of a many-body system can be characterized by three pairs of conjugate thermodynamic variables: $E/T,V/P,N/\mu$. In this limit, the thermodynamic properties in all ensembles are equivalent up to the leading order of $E,V,N$. However, for systems of finite size, this ensemble equivalence is no longer exact, and the thermodynamic properties may differ substantially among ensembles. To quantify these finite-size effects rigorously, it is desirable to develop a universal ensemble theory applicable to systems of arbitrary size, providing exact expressions for entropy and, thereby, giving rise to the precise value of all equilibrium thermodynamic quantities. In this work, we propose a theory that determines the exact entropy expressions for classical non-interacting many-body systems of arbitrary size across all statistical ensembles, based on only two postulates: \textbf{stationarity}, requiring that the physical laws be invariant under time translation, and \textbf{unbiasedness}, requiring that the equilibrium mixed state maximize the entropy subject to the prescribed constraints. Moreover, we show that the entropy expressions obtained in different ensembles converge to the common asymptotic form 
$S \asymp \ln\!\left(
\left( \frac{4\pi m e E}{3N} \right)^{3N/2} \cdot
\left(\frac{V}{N}\right)^N
\right)+N$,
consistent with the predictions of the large deviation theory.

\end{abstract}


{\noindent \footnotesize\textbf{*}Yuheng Wu,  \linkable{wuyuheng@mit.edu} }

\begin{spacing}{2}   

\section{Introduction}

Ensemble theory, initiated by Ludwig Boltzmann and James Clerk Maxwell and later formally proposed and developed by Josiah Willard Gibbs, is a well-established framework for formalizing equilibrium-state thermodynamics\cite{PathriaBeale2011}. In the thermodynamic limit, a system in thermal equilibrium can be specified by three independent extensive state variables $E,V,N$, or equivalently, by their conjugate intensive variables $T,P,\mu$, giving rise to eight ensembles. Accordingly to the large deviation theory, the thermodynamic properties in all eight ensembles are equivalent up to the leading order of $E,V,N$. However, beyond the thermodynamic limit, the thermodynamic properties of a small system can vary largely among different ensembles, and we still lack a generic framework that can precisely resolve these non-negligible differences.

Herein, in this work, we present a theory that provides a universal method for deriving the exact entropies of classical non-interacting many-body systems in arbitrary ensembles. With the exact entropy expression, all thermodynamic properties can be properly defined and precisely calculated. Notably, our framework does not invoke any approximation required in standard ensemble theory and, therefore, does not depend on the thermodynamic limit. The explicit expressions $S(E,V,N)$ for four representative ensembles take the following forms: 
\begin{align}
(E,V,N)\text{-Ensemble:}\quad
& \ln\!\left(
\frac{2\cdot\pi^{3N/2}(2mE)^{(3N-1)/2}}{\Gamma(3N/2)}
\cdot \frac{V^N}{N!}
\right),
\label{eq:EVN}
\\[1.2ex]
(\bar E,V,N)\text{-Ensemble:}\quad
& \ln\!\left(
\left( \frac{4\pi m e E}{3N} \right)^{3N/2} \cdot
\frac{V^N}{N!}
\right),
\label{eq:TVN}
\\[1.2ex]
(E,\bar V,N)\text{-Ensemble:}\quad
& \ln\!\left(
\frac{2\cdot\pi^{3N/2}(2mE)^{(3N-1)/2}}{\Gamma(3N/2)}
\right)
+(N+1)(1+\ln V-\ln (N+1)),
\label{eq:EPN}
\\[1.2ex]
(E,V,\bar N)\text{-Ensemble:}\quad
& \ln\!\left(
\sum_{n=1}^{\infty}
\frac{2\cdot\pi^{3n/2}(2mE)^{(3n-1)/2}}{\Gamma(3n/2)}\,\frac{V^n}{n!}\,e^{-\beta (n-N)} \right),
\label{eq:EVMu}
\end{align}
where $\beta$ is determined by $N=-\frac{\partial}{\partial \beta}\ln \left(\sum_{n=1}^\infty \frac{2\cdot\pi^{3n/2}(2mE)^{(3n-1)/2}}{\Gamma(3n/2)}\,\frac{V^n}{n!}\,e^{-\beta n}\right)$. It can be observed that the four representative ensembles each carry a distinct expression of entropy, representing four different ways to thermalize the system. But asymptotically, the four entropy expressions converge, up to the leading order of $E,V,N$, to the same limit
\begin{equation}
S \asymp \ln\!\left(
\left( \frac{4\pi m e E}{3N} \right)^{3N/2} \cdot
\left(\frac{V}{N}\right)^N
\right)+N,
\end{equation}
consistent with the predictions of the large deviation theory\cite{Touchette2009}. 

More broadly, the theory developed in this work demonstrates that equilibrium statistical mechanics extends well beyond its thermodynamic-limit approximation, motivating a re-examination of the regime of validity of conventional macroscopic thermodynamics. When the system size becomes sufficiently small, the asymptotic entropy expressions underlying macroscopic thermodynamics are no longer quantitatively accurate, and the exact finite-size entropy expressions must be employed to account for non-negligible finite-size corrections. A typical example is an ideal gas confined to nanoscale volumes, such as O$_2$ molecules trapped within nanopores. In such systems, finite-size effects can lead to appreciable deviations in temperature, pressure, and chemical potential from their bulk values. The present theory provides a rigorous framework for quantifying these deviations and thereby extends classical thermodynamics to finite-size systems.

\section{Thermal Equilibrium State of Classical Dynamical Systems}
The definition of thermal equilibrium states is the cornerstone of any theory of statistical mechanics. Generally, a self-consistent definition of the thermal equilibrium state begins with the Hamiltonian formulation of classical dynamical systems. In Hamiltonian mechanics, the status of the dynamical system exists in a symplectic manifold, namely phase space\cite{Arnold1989}. A \textit{pure state} is a state with complete information about the system, represented as a single point in phase space. A \textit{mixed state} is a probability distribution across multiple pure states, which carries uncertainty about the system's exact whereabouts. In statistical mechanics, we generalize the notion of a mixed state by allowing the statistical superposition of pure states drawn from phase spaces with different spatial volumes $V$ and particle numbers $N$. We refer to such a generalized probability distribution as a \emph{generalized mixed state}. The \emph{observables} in Hamiltonian mechanics are any smooth functions defined on the phase space of the dynamical system, and of time. Since Hamiltonian systems are driven by measure-preserving canonical transformations, the time evolution of mixed state densities and observables is well-defined. This allows the determination of thermal equilibrium states by two simple postulates: \emph{stationarity}, requiring the time evolution of the dynamical system to not depend on absolute time, and \emph{unbiasedness}, requiring the equilibrium mixed state to be subject to no bias that we have no evidence for. 

More precisely, stationarity means that the physical laws governing the dynamical system are invariant under time translations. Since every regular Hamiltonian system admits an invertible Legendre transformation to a non-degenerate Lagrangian system, we may invoke Noether's theorem to impose the stationarity condition. For a Lagrangian system, Noether's theorem states that the invariance of the Lagrangian under time translations implies the conservation of energy $E$ along physical trajectories. Therefore, for a pure state, stationarity requires its dynamical trajectory to remain at fixed $E,V,N$. For a generalized mixed state, which is a statistical mixture of pure states drawn from phase spaces with potentially different $V$ and $N$, stationarity instead requires the expectation values of $E,V,N$ to be conserved. Depending on our prior knowledge of these quantities, each of $E,V,N$ may be constrained either sharply to a fixed value or only through its expectation value, giving rise to eight distinct ensembles. For example, when $E,V,N$ are all sharply fixed, the resulting ensemble is the microcanonical ensemble, which we denote by $(E,V,N)$-Ensemble. When $E$ is constrained only in mean while $V$ and $N$ are sharply fixed, the resulting ensemble is the canonical ensemble, denoted by $(\bar E,V,N)$-Ensemble. These eight ensembles may also be interpreted as corresponding to eight distinct mechanisms of thermalization. A typical example is Langevin dynamics, which models thermalization through stochastic differential equations containing both dissipative and fluctuating forces. Under appropriate fluctuation-dissipation conditions, the resulting stochastic dynamics admits the canonical distribution as its stationary distribution, corresponding precisely to the equilibrium mixed state of $(\bar E,V,N)$-Ensemble\cite{Shiraishi2023}.

The unbiasedness condition is formulated through the principle of maximum entropy\cite{Jaynes1957}. In statistical inference, among all probability distributions consistent with a given set of constraints, the distribution that maximizes the entropy represents the least biased assignment of probabilities\cite{CoverThomas2006}. Accordingly, in statistical mechanics, the most unbiased generalized mixed state among all admissible mixed states is the one that maximizes the entropy subject to the prescribed constraints.

Based on these two postulates, the equilibrium mixed state can be determined accordingly. In what follows, we restrict our analysis to the simplest class of regular Hamiltonian systems, namely classical non-interacting many-body systems. We further require that the constituent particles be identical and impose \emph{indistinguishability} as a symmetry of the many-body state space. Formally, indistinguishability means a permutation symmetry on phase space\cite{Leinaas1977}. Let $P_N$ denotes the \textit{permutation group} of $N$ elements; its action on the $N$-particle phase space $\Omega_N$ is given by
\begin{equation}
\pi \cdot (q_1,p_1,\dots,q_N,p_N)
=
(q_{\pi(1)},p_{\pi(1)},\dots,q_{\pi(N)},p_{\pi(N)}),
\qquad \pi \in P_N .
\end{equation}
The physical phase space is not the labeled space $\Omega_N$, but its quotient $\widetilde{\Omega}_N = \Omega_N / P_N$, whose points are equivalence classes $[(\{q_i\},\{p_i\})]
=
\{\pi \cdot z \mid \pi \in P_N\}$. And the physical pure state is no longer a single phase-space point, but an entire \textit{permutation orbit}. Accordingly, mixed states are described by a probability density $\widetilde{\rho}(\{q_i\},\{p_i\})$ defined on the quotient space $\widetilde{\Omega}_N$, with an invariant measure $d\widetilde{\Omega}
=
\frac{1}{N!}\,d\Omega$. The extra $N!$ term compensates for the overcounting of physically identical configurations in $\Omega_N$, ensuring that ensemble averages are defined on the true space of physical states and providing the classical resolution of the Gibbs paradox. Technically speaking, the true indistinguishability correction term, in a more precise sense, is $\frac{N!}{\prod_\alpha n_\alpha!}$, where \(n_\alpha\) denotes the occupation number of the one-particle state \(\alpha\), and $\prod_\alpha n_\alpha!$ is the order of the stabilizer subgroup consisting of permutations that leave the configuration invariant. But for classical dynamical systems whose phase spaces are continuous, the stabilizer is almost surely $1$, because the particle configurations with phase-space overlap span a set of Liouville measure zero. In other words, a countable permutation orbit generically contains no coincident particles, and thereby no nontrivial stabilizer arises.

\section{Exact Expressions of Entropy}
\begin{figure}[H]
    \centering
    \includegraphics[width=0.9\linewidth]{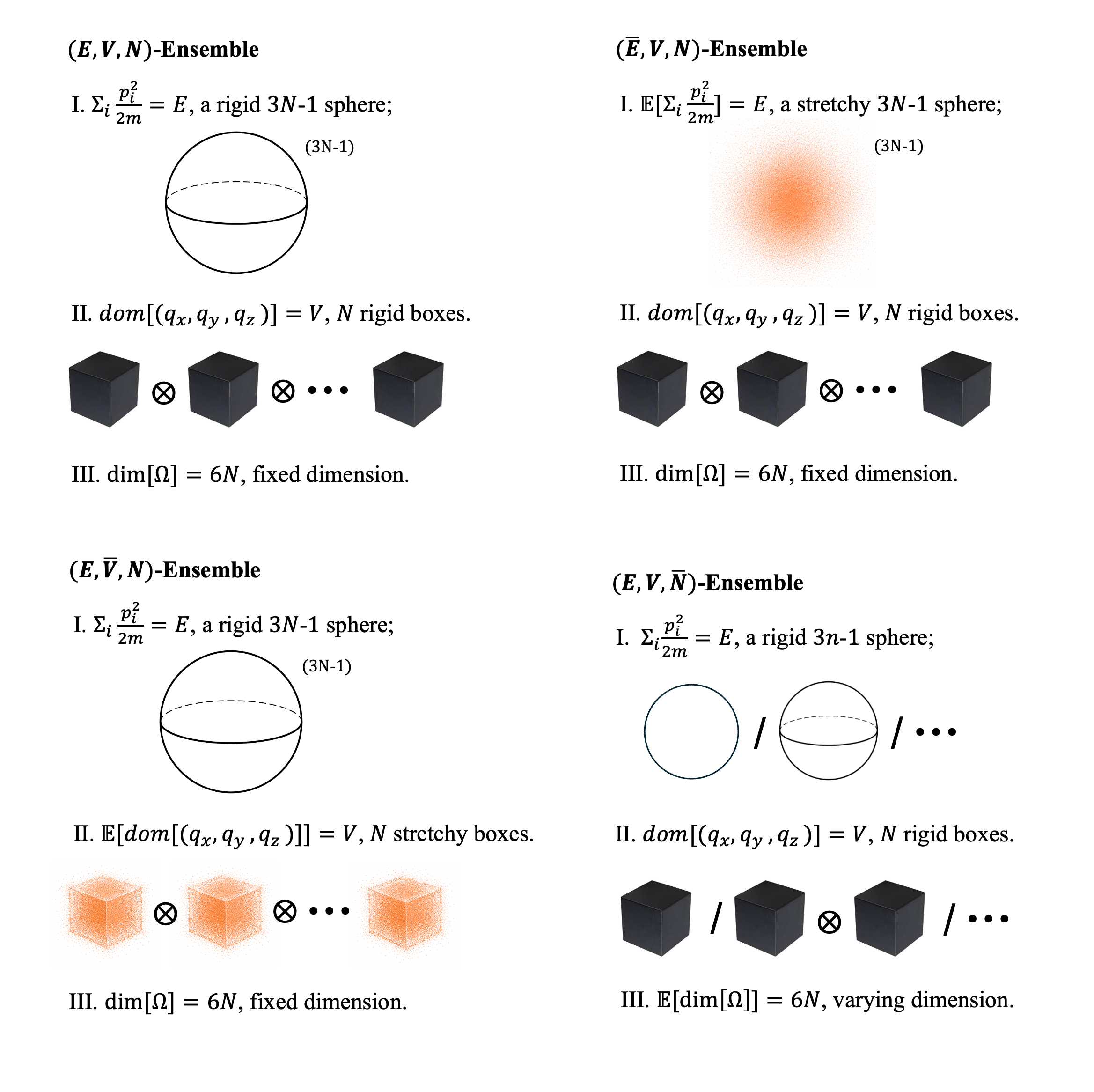}
    \vspace{1em}
    \caption{Schematics of the state space and constraints for the four representative ensembles.}
    \label{ensemble-schematics}
\end{figure}

Once the Lagrangian of a dynamical system is specified, the equilibrium mixed state in each ensemble is determined. For classical non-interacting many-body systems, the Lagrangian takes a simple form $\mathcal L=\sum_{i=1}^{3N}\frac{p_i^2}{2m}$, therefore, $E=\sum_{i=1}^{3N}\frac{p_i^2}{2m}$. By maximizing the entropy in each ensemble, the exact expression of entropy can be obtained accordingly (Fig.~\ref{ensemble-schematics}). Without loss of generality, we restrict our following discussion on four typical ensembles, $(E,V,N)$-Ensemble, $(\bar E,V,N)$-Ensemble, $(E,\bar V,N)$-Ensemble, and $(E,V,\bar N)$-Ensemble, because the conclusions of the other four ensembles become trivial once these four are done.

\subsection{$(E,V,N)$-Ensemble}

We begin with the $(E,V,N)$-Ensemble, the simplest realization of thermal equilibrium for a dynamical system. In this ensemble, the energy $E$, the accessible spatial volume $V$, and the particle number $N$ are set to definite values. Before imposing particle indistinguishability, the resulting state space is the product of two parts: a product space of $N$ rigid hypercubes of volume $V$ for spatial coordinates $\{q_i\}$, and a rigid hypersphere of dimension $3N-1$ for momentum coordinates $\{p_i\}$, denote as $\Lambda_{E,V,N}$. Obviously, since the state space volume of $(E,V,N)$-Ensemble is finite, the maximum-entropy optimization simply yields uniform probability density over state space. The standard derivation invokes the Lagrange multiplier method with a single multiplier enforcing normalization:
\begin{equation}
\mathcal{L}[\rho]
=
-\int_{\Lambda_{E,V,N}}\rho\ln\rho\,d\Omega
-\alpha\!\left(
\int_{\Lambda_{E,V,N}}\rho\,d\Omega-1
\right).
\end{equation}
Taking a first variation \(\rho\mapsto\rho+\delta\rho\) yields
\begin{equation}
\delta\mathcal{L}
=
-\int_{\Lambda_{E,V,N}}
\delta\rho\,
\left(
\ln\rho+1+\alpha
\right)
\,d\Omega.
\end{equation}
Since \(\delta\rho\) is arbitrary, stationarity implies
\begin{equation}
\begin{aligned}
&\ln\rho(\{q_i\},\{p_i\})+1+\alpha=0,\\
&\rho(\{q_i\},\{p_i\})
=
e^{-1-\alpha}.
\end{aligned}
\label{EVN_family}
\end{equation}
Normalization requires
\begin{equation}
\begin{aligned}
1
&=
\int_{\Lambda_{E,V,N}}
\rho(\{q_i\},\{p_i\})
\,d\Omega
=
e^{-1-\alpha}
\int_{\Lambda_{E,V,N}}d\Omega\\
&=
e^{-1-\alpha}\;\mathrm{Surf}_{3N-1}(\sqrt{2mE})\,V^N\\
&=
e^{-1-\alpha}\cdot \frac{
2\cdot\pi^{3N/2}(2mE)^{(3N-1)/2}
}{
\Gamma(3N/2)
}
\,V^N.
\end{aligned}
\end{equation}
Therefore
\begin{equation}
e^{-1-\alpha}=\frac{
2\cdot\pi^{3N/2}(2mE)^{(3N-1)/2}
}{
\Gamma(3N/2)
}
\,V^N
\equiv
\frac{1}{Z},
\end{equation}
where $Z$, the normalization factor, is known as the partition function. Therefore, the probability density function of the equilibrium mixed state and the associated entropy in $(E,V,N)$-Ensemble are given by
\begin{equation}
\begin{aligned}
&\rho^\mathrm{eq}_{_{E,V,N}}(\{q_i\},\{p_i\})
= \frac{\Gamma(3N/2)}{2\cdot\pi^{3N/2}(2mE)^{(3N-1)/2}}
\cdot \frac{1}{V^N}, \\
&S[\rho^\mathrm{eq}_{_{E,V,N}}]
=
\ln\!\left(
\frac{2\cdot\pi^{3N/2}(2mE)^{(3N-1)/2}}{\Gamma(3N/2)}
\cdot V^N
\right).
\end{aligned}
\end{equation}
After incorporating the indistinguishability of particles, the true physical equilibrium mixed state and entropy take the form of
\begin{equation}
\begin{aligned}
&\tilde\rho^\mathrm{eq}_{_{E,V,N}}
= \frac{\Gamma(3N/2)}{2\cdot\pi^{3N/2}(2mE)^{(3N-1)/2}}
\cdot \frac{N!}{V^N}, \\
&S[\tilde\rho^\mathrm{eq}_{_{E,V,N}}]
=
\ln\!\left(
\frac{2\cdot\pi^{3N/2}(2mE)^{(3N-1)/2}}{\Gamma(3N/2)}
\cdot \frac{V^N}{N!}
\right).
\end{aligned}
\end{equation}

\subsection{$(\bar{E},V,N)$-Ensemble} 

We next consider the $(\bar{E},V,N)$-Ensemble, in which volume $V$ and particle number $N$ remain sharply fixed while the total energy $E$ is constrained only in mean. In this case, the spatial part of the state space still consists of $N$ rigid hypercubes, while the momentum part is now a stretchable hypersphere that explores the full $\mathbb{R}^{3N}$ subject only to the mean-energy constraint. We denote this state space as $\Lambda_{\bar E,V,N}$. To determine the equilibrium mixed state, we consider the following Lagrangian:
\begin{equation}
\mathcal{L}[\rho]
=
-\int_{\Lambda_{\bar E,V,N}}\rho\ln\rho\,d\Omega
-\alpha\!\left(\int_{\Lambda_{\bar E,V,N}}\rho\,d\Omega-1\right)
-\beta\!\left(\int_{\Lambda_{\bar E,V,N}}\rho\cdot\sum_{i=1}^{3N}\frac{p_i^2}{2m}\,d\Omega-E\right).
\end{equation}
Taking a first variation $\rho\mapsto\rho+\delta\rho$ yields
\begin{equation}
\delta\mathcal{L}
=
-\int_{\Lambda_{\bar E,V,N}}\delta\rho\cdot\left(\ln\rho+1+\alpha+\beta\cdot\sum_{i=1}^{3N}\frac{p_i^2}{2m}\right)\,d\Omega.
\end{equation}
Since $\delta\rho$ is arbitrary, stationarity implies
\begin{equation}
\begin{aligned}
&\ln\rho(\{q_i\},\{p_i\}) + 1 + \alpha + \beta \cdot \sum_{i=1}^{3N}\frac{p_i^2}{2m}=0,\\
&\rho(\{q_i\},\{p_i\})
=
e^{-1-\alpha}\,e^{-\beta \sum_{i=1}^{3N}\frac{p_i^2}{2m}}.
\end{aligned}
\label{barEVN_family}
\end{equation}
Normalization requires
\begin{equation}
\begin{aligned}
1=\int_{\Lambda_{\bar E,V,N}}\rho(\{q_i\},\{p_i\})\,d\Omega
&=e^{-1-\alpha}\cdot\int_{\Lambda_{\bar E,V,N}}\prod _{i=1}^{3N}e^{-\beta \frac{p_i^2}{2m}}\,d\Omega
=e^{-1-\alpha}\cdot\,V^N\cdot \left(\frac{2\pi m}{\beta}\right)^{\frac{3N}{2}},\\
\frac{1}{Z(\beta)}&\equiv e^{-1-\alpha}=\frac{1}{V^N}\left(\frac{\beta}{2\pi m}\right)^{\frac{3N}{2}}.
\end{aligned}
\end{equation}
The constraint of mean energy enforces
\begin{equation}
E=\int_{\Lambda_{\bar E,V,N}}\sum_{i=1}^{3N}\frac{p_i^2}{2m}\cdot\rho(\{q_i\},\{p_i\})\,d\Omega=\frac{1}{Z(\beta)}\cdot\int_{\Lambda_{\bar E,V,N}}\sum_{i=1}^{3N}\frac{p_i^2}{2m}\cdot e^{-\beta \frac{p_i^2}{2m}}\,d\Omega=\frac{3N}{2\beta}.
\end{equation}
The equilibrium mixed state is therefore
\begin{equation}
\rho_{_{\bar E,V,N}}^{\mathrm{eq}}(\{q_i\},\{p_i\})
=
\frac{1}{V^N}\left(\frac{3N}{4\pi mE}\right)^{\frac{3N}{2}}
\cdot
\exp\!\left(-\sum_{i=1}^{3N}\frac{3N\cdot p_i^2}{4mE}\right).
\label{rho_eq_T}
\end{equation}
Since a Gaussian random variable $\mathcal{N}(\mu,\sigma^{2})$ has differential entropy $\frac{1}{2}\ln(2\pi e\,\sigma^{2})$, it follows that
\begin{equation}
S[\rho_{_{\bar E,V,N}}^{\mathrm{eq}}]
=
 \ln\!\left(
\left(\frac{4\pi m e E}{3N}\right)^{3N/2} \cdot\frac{V^N}{N!}
\right).
\end{equation}
Taking into account indistinguishability, the true physical equilibrium mixed state and entropy of $(\bar E,V,N)$-Ensemble are
\begin{equation}
\begin{aligned}
&\tilde\rho_{_{\bar E,V,N}}^{\mathrm{eq}}
=
\frac{N!}{V^N}\left(\frac{3N}{4\pi mE}\right)^{\frac{3N}{2}}
\cdot
\exp\!\left(-\sum_{i=1}^{3N}\frac{3N\cdot p_i^2}{4mE}\right).\\
&S[\tilde\rho_{_{\bar E,V,N}}^{\mathrm{eq}}]
=
\ln\!\left(
\left(\frac{4\pi m e E}{3N}\right)^{3N/2} \cdot\frac{V^N}{N!}
\right).
\end{aligned}
\end{equation}

\subsection{$(E,\bar{V},N)$-Ensemble}
Next, we consider the $({E},\bar V,N)$-Ensemble, where energy $E$ and particle number $N$ are sharply fixed while the volume $V$ is constrained in mean. The state space is now a rigid $(3N-1)$-dimensional rigid hypersphere for momentum coordinates and the product space of $N$ stretchable cubes of fixed mean volume for spatial coordinates. We denote this state space as $\Lambda_{E,\bar V,N}$, and the observable, the volume of the spatial domain, as $\nu$. Unlike \((\bar E,V,N)\)-Ensemble, where the energy observable $\varepsilon
=
\sum_{i=1}^{3N}\frac{p_i^2}{2m}$ is a well-defined observable on the phase space spanned by $\{q_i\},\{p_i\}$, the volume observable \(\nu\) is not a deterministic function of the phase space coordinates $\{q_i\},\{p_i\}$. In fact, the admissible mixed states in $({E},\bar V,N)$-Ensemble live in an extended state space parametrized by \((\{q_i\},\{p_i\},\nu)\), equipped with an invariant measure $d\Omega
=
d\{q_i\}\,d\{p_i\}\,d\nu$. And the extended state space \(\Lambda_{E,\bar V,N}\) admits a decomposition 
\begin{equation}
\Lambda_{E,\bar V,N}
=
\bigsqcup_{\nu}
\Lambda(\nu), 
\end{equation}
where \(\Lambda(\nu)\subseteq\Lambda_{E,\bar V,N}\) denotes the phase space corresponding to a fixed spatial volume \(\nu\). In other words, \(\Lambda_{E,\bar V,N}\) is the disjoint union of all possible phase spaces with different spatial volumes that have the same energy $E$ and particle number $N$. To determine the maximum-entropy mixed state, we consider the Lagrangian
\begin{equation}
\mathcal{L}[\rho]
=
-\int_{\Lambda_{E,\bar V,N}}
\rho\ln\rho\,d\Omega
-\alpha
\left(
\int_{\Lambda_{E,\bar V,N}}
\rho\,d\Omega
-1
\right)
-\beta
\left(
\int_{\Lambda_{E,\bar V,N}}
\nu\,\rho\,d\Omega
- V
\right).
\end{equation}
Taking a first variation $\rho\mapsto\rho+\delta\rho$ yields
\begin{equation}
\delta\mathcal{L}
=
-\int_{\Lambda_{ E,
\bar V,N}}\delta\rho\cdot\left(\ln\rho+1+\alpha+\beta\nu\right)\,d\Omega.
\end{equation}
Since $\delta\rho$ is arbitrary, stationarity implies
\begin{equation}
\begin{aligned}
&\ln\rho(\{q_i\},\{p_i\},\nu) + 1 + \alpha + \beta \nu=0,\\
&\rho(\{q_i\},\{p_i\},\nu)
=
e^{-1-\alpha}\,e^{-\beta \nu}.
\end{aligned}
\label{barEVN_family}
\end{equation}
Factorizing $\rho$ into conditional and marginal parts yields
\begin{equation}
    \rho\!\left(\{q_i\},\{p_i\}, \nu\right)=\rho\!\left(\{q_i\},\{p_i\}\mid \nu\right)\cdot\rho(\nu).
    \label{factorize}
\end{equation}
Since \eqref{barEVN_family} suggests that the equilibrium mixed-state density has no explicit dependence on $\{q_i\},\{p_i\}$, its conditional part, $\rho\!\left(\{q_i\},\{p_i\}\mid \nu\right)$, must also be independent of $\{q_i\},\{p_i\}$, i.e.
\begin{equation}
\rho(\{q_i\},\{p_i\}\mid \nu)
=
\frac{\Gamma(3N/2)}{2\cdot\pi^{3N/2}(2mE)^{(3N-1)/2}}
\cdot \frac{1}{\nu^N}.
\label{eq:EVN_max_nu}
\end{equation}
To determine the marginal part $\rho(\nu)$, we insert \eqref{factorize} into the Lagrangian, which gives
\begin{equation}
\begin{aligned}
\mathcal{L}[\rho]
=& - \int_{0}^{\infty} \rho(\nu)
\left(
\int_{\Lambda(\nu)}
\rho\!\left(\{q_i\},\{p_i\}\mid \nu\right)
\ln \rho\!\left(\{q_i\},\{p_i\}\mid \nu\right)
\, d\{q_i\}\,d\{p_i\}
\right) d\nu\\
&-\;
\int_{0}^{\infty} \rho(\nu)\ln\rho(\nu)\, d\nu 
\\
&
- \alpha\left(\int_{0}^{\infty} \rho(\nu)
\left(
\int_{\Lambda(\nu)}
\rho\!\left(\{q_i\},\{p_i\}\mid \nu\right)
\, d\{q_i\}\,d\{p_i\}
\right) d\nu-1\right)
\\
&
- \beta\left(\int_{0}^{\infty} \nu\rho(\nu)
\left(
\int_{\Lambda(\nu)}
\rho\!\left(\{q_i\},\{p_i\}\mid \nu\right)
\, d\{q_i\}\,d\{p_i\}
\right) d\nu-V\right).
\end{aligned}
\end{equation}
Substituting \eqref{eq:EVN_max_nu} into the Lagrangian yields
\begin{equation}
\begin{aligned}
\mathcal{L}[\rho]
=
&\ln\!\left(
\frac{2\cdot\pi^{3N/2}(2mE)^{(3N-1)/2}}{\Gamma(3N/2)}
\right) +
\int_{0}^{\infty}\rho(\nu)\,\ln\frac{\nu^{N}}{\rho(\nu)}\,d\nu
-\alpha\!\left(\int_{0}^{\infty}\rho(\nu)\,d\nu-1\right)
\\&
-\beta\!\left(\int_{0}^{\infty}\nu\,\rho(\nu)\,d\nu-V\right).
\end{aligned}
\label{eq:Lagrangian_functional}
\end{equation}
Taking a first variation $\rho\mapsto\rho+\delta\rho$ yields
\begin{equation}
\delta\mathcal{L}
=
\int_{0}^{\infty}\delta\rho\cdot\Bigl(\ln\frac{\nu^{N}}{\rho(\nu)}-1-\alpha-\beta \nu\Bigr)\,d\nu.
\label{eq:first_variation_total}
\end{equation}
Stationarity requires $\delta\mathcal{L}=0$ for arbitrary variations $\delta\rho$, yielding
\begin{equation}
\rho(\nu)=e^{-1-\alpha}\,\nu^{N}e^{-\beta \nu}.
\label{eq:rho_solution_form}
\end{equation}
Normalization requires
\begin{equation}
\begin{aligned}
1=\int_{0}^{\infty}\rho(\nu)\,d\nu
&=e^{-1-\alpha}\cdot\int_{0}^{\infty}\nu^{N}e^{-\beta \nu}\,d\nu
=e^{-1-\alpha}\cdot\,\frac{\Gamma(N+1)}{\beta^{N+1}},\\
\frac{1}{Z(\beta)}&\equiv e^{-1-\alpha}=\frac{\beta^{N+1}}{\Gamma(N+1)}.
\end{aligned}
\end{equation}
The constraint of fixed mean volume requires
\begin{equation}
V=\int_{0}^{\infty}\nu\,\rho(\nu)\,d\nu
=\frac{1}{Z(\beta)}\cdot\int_{0}^{\infty}\nu^{N+1}e^{-\beta \nu}\,d\nu
=\frac{1}{Z(\beta)}\cdot\,\frac{\Gamma(N+2)}{\beta^{N+2}}
=\frac{N+1}{\beta}.
\end{equation}
Together, the marginal part $\rho(\nu)$ is a Gamma distribution with shape $N+1$ and rate $(N+1)/V$:
\begin{equation}
\rho(\nu)
=
\frac{\bigl(\frac{N+1}{V}\bigr)^{N+1}}{\Gamma(N+1)}\cdot
\nu^{N}\cdot\exp\!\left(-\frac{N+1}{V}\,\nu\right),
\label{eq:gamma_distribution}
\end{equation}
i.e. $\nu\sim\Gamma\!\left(N+1,(N+1)/V\right)$. The unconditional equilibrium mixed state density is therefore
\begin{equation}
\rho_{_{E,\bar V,N}}^{\mathrm{eq}}(\{q_i\},\{p_i\},\nu)=\frac{\Gamma(3N/2)}{2\cdot\pi^{3N/2}(2mE)^{(3N-1)/2}}
\cdot \frac{\bigl(\frac{N+1}{V}\bigr)^{N+1}}{\Gamma(N+1)}
\cdot\exp\!\left(-\frac{N+1}{V}\,\nu\right).
\end{equation}
Substituting it into the Lagrangian, we can obtain the expression of entropy in $(E,\bar V, N)$-Ensemble:
\begin{equation}
S[\rho_{_{E,\bar V,N}}^{\mathrm{eq}}]
=
\ln\!\left(
\frac{2\cdot\pi^{3N/2}(2mE)^{(3N-1)/2}}{\Gamma(3N/2)}
\right)
+(N+1)(1+\ln V-\ln (N+1))+\ln N!.
\end{equation}
After imposing the indistinguishability of particles, the physical mixed-state and entropy become
\begin{equation}
\begin{aligned}
& \tilde\rho_{_{E,\bar V,N}}^{\mathrm{eq}}=\frac{\Gamma(3N/2)}{2\cdot\pi^{3N/2}(2mE)^{(3N-1)/2}}
\cdot \bigl(\frac{N+1}{V}\bigr)^{N+1}
\cdot\exp\!\left(-\frac{N+1}{V}\,\nu\right).\\
& S[\rho_{_{E,\bar V,N}}^{\mathrm{eq}}]
=
\ln\!\left(
\frac{2\cdot\pi^{3N/2}(2mE)^{(3N-1)/2}}{\Gamma(3N/2)}
\right)
+(N+1)(1+\ln V-\ln (N+1)).
\end{aligned}
\end{equation}

\subsection{$(E,V,\bar{N})$-Ensemble}
Finally, we consider the $({E}, V,\bar N)$-Ensemble, where energy $E$ and volume $V$ are sharply fixed, but particle number $N$ is constrained in mean. Analogous to $({E},\bar V,N)$-Ensemble, the extended state space for $({E}, V,\bar N)$-Ensemble is the disjoint union of all possible phase spaces with different particle numbers that have the same energy $E$ and spatial volume $V$:
\begin{equation}
\Lambda_{E,V,\bar N}
=
\bigsqcup_{n}
\Lambda(n),
\end{equation}
equipped with an invariant measure $d\Omega
=
d\{q_i\}\,d\{p_i\}\,d\mu_{\mathrm{count}}(n)$, where $d\mu_{\mathrm{count}}(n)$ denotes the counting measure on $n\in\mathbb{N}^+$. To determine the maximum-entropy mixed state, we consider the Lagrangian
\begin{equation}
\mathcal{L}[\rho]
=
-\int_{\Lambda_{E,V,\bar N}}
\rho\ln\rho\,d\Omega
-\alpha
\left(
\int_{\Lambda_{E,V,\bar N}}
\rho\,d\Omega
-1
\right)
-\beta
\left(
\int_{\Lambda_{E,V,\bar N}}
n\,\rho\,d\Omega
- N
\right).
\end{equation}
Taking a first variation $\rho\mapsto\rho+\delta\rho$ yields
\begin{equation}
\delta\mathcal{L}
=
-\int_{\Lambda_{ E,
V,\bar N}}\delta\rho\cdot\left(\ln\rho+1+\alpha+\beta n\right)\,d\Omega.
\end{equation}
Since $\delta\rho$ is arbitrary, stationarity implies
\begin{equation}
\begin{aligned}
&\ln\rho(\{q_i\},\{p_i\},n) + 1 + \alpha + \beta n=0,\\
&\rho(\{q_i\},\{p_i\},n)
=
e^{-1-\alpha}\,e^{-\beta n}.
\end{aligned}
\label{EVbarN_family}
\end{equation}
Factorizing $\rho$ into conditional and marginal parts yields
\begin{equation}
\rho\!\left(\{q_i\},\{p_i\}, n\right)=\rho\!\left(\{q_i\},\{p_i\}\mid n\right)\cdot\rho(n),
\end{equation}
the conditional part, $\rho\!\left(\{q_i\},\{p_i\}\mid n\right)$, must be independent of $\{q_i\},\{p_i\}$, because \eqref{EVbarN_family} suggests that the equilibrium mixed-state density has no explicit dependence on $\{q_i\},\{p_i\}$. Therefore,
\begin{equation}
\rho(\{q_i\},\{p_i\}\mid n)
=
\frac{\Gamma(3n/2)}{2\cdot\pi^{3n/2}(2mE)^{(3n-1)/2}}
\cdot \frac{1}{V^n}.
\label{conditional_n}
\end{equation}
To determine the marginal part $\rho(n)$, we factorize the $\rho(\{q_i\},\{p_i\},n)$ in the Lagrangian:
\begin{equation}
\begin{aligned}
\mathcal{L}[\rho]
=& - \sum_{n=1}^{\infty} \rho(n)
\left(
\int_{\Lambda(n)}
\rho\!\left(\{q_i\},\{p_i\}\mid n\right)
\ln \rho\!\left(\{q_i\},\{p_i\}\mid n\right)
\, d\{q_i\}\,d\{p_i\}
\right)\\
&-\;
\sum_{n=1}^{\infty} \rho(n)\ln\rho(n)
\\
&
- \alpha\left(\sum_{n=1}^{\infty} \rho(n)
\left(
\int_{\Lambda(n)}
\rho\!\left(\{q_i\},\{p_i\}\mid n\right)
\, d\{q_i\}\,d\{p_i\}
\right)-1\right)
\\
&
- \beta\left(\sum_{n=1}^{\infty} n\rho(n)
\left(
\int_{\Lambda(n)}
\rho\!\left(\{q_i\},\{p_i\}\mid n\right)
\, d\{q_i\}\,d\{p_i\}
\right)-N\right).
\end{aligned}
\label{Lag_n_factor}
\end{equation}
Inserting \eqref{conditional_n} into the \eqref{Lag_n_factor} yields
\begin{equation}
\begin{aligned}
\mathcal{L}[\rho]
=&
\sum_{n=1}^{\infty}\rho(n)\ln \left( \frac{2\cdot\pi^{3n/2}(2mE)^{(3n-1)/2}}{\Gamma(3n/2)}\,V^n \right)
-\sum_{n=1}^{\infty}\rho(n)\ln\rho(n)
-\alpha\Bigl(\sum_{n=1}^{\infty}\rho(n)-1\Bigr)\\&
-\beta\Bigl(\sum_{n=1}^{\infty}n\,\rho(n)-N\Bigr).
\end{aligned}
\end{equation}
For notational convenience, we introduce
\begin{equation}
\mathcal{G}(n)
\equiv
\frac{2\cdot\pi^{3n/2}(2mE)^{(3n-1)/2}}{\Gamma(3n/2)}\,V^n.
\label{eq:Sn_def}
\end{equation}
Taking a first variation $\rho\mapsto\rho+\delta\rho$ yields
\begin{align}
\delta\mathcal{L}
&=
\sum_{n=1}^{\infty}\delta\rho(n)\ln \mathcal{G}(n)
-\sum_{n=1}^{\infty}\delta\rho(n)\bigl(\ln\rho(n)+1\bigr)
-\alpha\sum_{n=1}^{\infty}\delta\rho(n)
-\beta\sum_{n=1}^{\infty}n\,\delta\rho(n)
\nonumber\\
&=
\sum_{n=1}^{\infty}\delta\rho(n)
\Bigl(\ln \mathcal{G}(n)-\ln\rho(n)-1-\alpha-\beta n\Bigr).
\label{eq:n_first_variation}
\end{align}
$\delta\mathcal{L}=0$ for arbitrary variations $\delta\rho(n)$ requires
\begin{equation}
\begin{aligned}
&\ln \mathcal{G}(n)-\ln\rho(n)-1-\alpha-\beta n=0,\quad \rho(n)=e^{-1-\alpha}\,\mathcal{G}(n)e^{-\beta n}.
\end{aligned}
\label{eq:n_solution_form}
\end{equation}
Normalization requires
\begin{equation}
\begin{aligned}
&1=\sum_{n=1}^{\infty}\rho(n)
=e^{-1-\alpha}\sum_{n=1}^{\infty}\mathcal{G}(n)e^{-\beta n},\\
& \frac{1}{Z(\beta)}\equiv e^{-1-\alpha}=\frac{1}{\sum_{n=1}^{\infty}\mathcal{G}(n)e^{-\beta n}}.
\end{aligned}
\end{equation}
The mean particle number constraint requires
\begin{equation}
N=\sum_{n=1}^{\infty}n\,\rho(n)
=-\frac{\partial}{\partial \beta}\ln Z(\beta).
\label{eq:n_beta_constraint}
\end{equation}
Consequently, the entropy-maximizing marginal is
\begin{equation}
\rho(n)=\frac{1}{Z(\beta)}\,
\frac{2\cdot\pi^{3n/2}(2mE)^{(3n-1)/2}}{\Gamma(3n/2)}\,V^n\,e^{-\beta n}.
\end{equation}
The unconditional maximum-entropy mixed state of $(E,V,\bar N)$-Ensemble is
\begin{equation}
\rho_{_{E,V,\bar N}}^{\mathrm{eq}}\left(\{q_i\},\{p_i\}, n\right)=\frac{\rho(n)}{\mathcal{G}(n)} =
\frac{e^{-\beta n}}{Z(\beta)}=\frac{e^{-\beta n}}{\sum_{n=1}^{\infty}\frac{2\cdot\pi^{3n/2}(2mE)^{(3n-1)/2}}{\Gamma(3n/2)}\,V^ne^{-\beta n}},
\label{EVbarN_mixed}
\end{equation}
where $\beta$ is a constant that can be uniquely determined using \eqref{eq:n_beta_constraint}. Substituting \eqref{EVbarN_mixed} into the Lagrangian gives the exact expression of entropy
\begin{equation}
S[\rho_{_{E,V,\bar N}}^{\mathrm{eq}}]
=
\ln Z(\beta)+\beta N
=
\ln\!\left(
\sum_{n=1}^{\infty}
\frac{2\cdot\pi^{3n/2}(2mE)^{(3n-1)/2}}{\Gamma(3n/2)}\,V^n\,e^{-\beta (n-N)}
\right).
\end{equation}
Further imposing particle indistinguishability, the physical mixed-state density and entropy are given by
\begin{equation}
\begin{aligned}
&
\tilde\rho_{_{E,V,\bar N}}^{\mathrm{eq}}
=\frac{e^{-\beta n}}{\sum_{n=1}^{\infty}\frac{2\cdot\pi^{3n/2}(2mE)^{(3n-1)/2}}{\Gamma(3n/2)}\,\frac{V^n}{n!}e^{-\beta n}},\\
&
S[\tilde\rho_{_{E,V,\bar N}}^{\mathrm{eq}}]
=
\ln Z(\beta)+\beta N
=
\ln\!\left(
\sum_{n=1}^{\infty}
\frac{2\cdot\pi^{3n/2}(2mE)^{(3n-1)/2}}{\Gamma(3n/2)}\,\frac{V^n}{n!}\,e^{-\beta (n-N)}
\right).
\end{aligned}
\end{equation}

\section{Asymptotic Analysis}
So far, we have derived the exact entropy expressions for classical non-interacting many-body systems. To complete the theory, it remains to verify that the entropy expressions of the different ensembles become equivalent in the thermodynamic limit, thereby recovering the ensemble equivalence required by conventional thermodynamics. More precisely, we want to show that for every
$\xi,\zeta\in(0,\infty)$, along the thermodynamic scaling $E=\xi N,\,V=\zeta N,\,N\to\infty$, the entropy expressions in all ensembles asymptotically converge to the same limit
\begin{equation}
S \asymp \ln\!\left(
\left( \frac{4\pi m e E}{3N} \right)^{3N/2} \cdot
\left(\frac{V}{N}\right)^N
\right)+N.
\label{asym_limit}
\end{equation}
Without loss of generality, we focus on the four representative entropy expressions derived above, since the remaining four ensembles can be analyzed by similar methods. 

\subsection{$(E,V,N)$-Ensemble}

For the $(E,V,N)$-ensemble,
\begin{equation}
S[\tilde\rho_{_{E,V,N}}^{\mathrm{eq}}]
=
\ln\!\left(
\frac{
2\cdot\pi^{3N/2}
(2mE)^{(3N-1)/2}
}{
\Gamma(3N/2)
}
\frac{V^N}{N!}
\right).
\end{equation}
Using Stirling's formula,
\begin{equation}
\begin{aligned}
&
\lim_{x\to \infty}\Gamma(x)
=
\sqrt{2\pi}\,
x^{x-\frac12}e^{-x}
(1+o(1)),\\
&
\lim_{N\to \infty}N!
=
\sqrt{2\pi N}
\left(\frac{N}{e}\right)^N
(1+o(1)),
\label{Stirling}
\end{aligned}
\end{equation}
we obtain
\begin{equation}
\begin{aligned}
\lim_{E,V,N\to\infty}S[\tilde\rho_{_{E,V,N}}^{\mathrm{eq}}]
&= \ln\!
\left(
\left( \frac{4\pi m e E}{3N} \right)^{3N/2} \cdot
\left(\frac{V}{N}\right)^N
\right)+N
+\frac12
\ln
\left(
\frac{3}{4\pi^2 mE}
\right)
+o(\ln N)\\
&
= \ln\!
\left(
\left( \frac{4\pi m e E}{3N} \right)^{3N/2} \cdot
\left(\frac{V}{N}\right)^N
\right)+N
+O(\ln N).
\end{aligned}
\end{equation}
Therefore,
\begin{equation}
S[\tilde\rho_{_{E,V,N}}^{\mathrm{eq}}]
\asymp \ln\!
\left(
\left( \frac{4\pi m e E}{3N} \right)^{3N/2} \cdot
\left(\frac{V}{N}\right)^N
\right)+N.
\end{equation}

\subsection{$(\bar E,V,N)$-Ensemble}
For $(\bar E,V,N)$-Ensemble, the asymptotic analysis is even more obvious. Using \eqref{Stirling},
\begin{equation}
\begin{aligned}
\lim_{E,V,N\to \infty} S[\tilde\rho_{_{\bar E,V,N}}^{\mathrm{eq}}]
& =
\lim_{E,V,N\to \infty}\ln\!\left(
\left(\frac{4\pi m e E}{3N}\right)^{3N/2} \cdot\frac{V^N}{N!}
\right)\\ & =
\ln\!\left(
\left(\frac{4\pi m e E}{3N}\right)^{3N/2} \cdot\left(\frac{V}{N}\right)^N
\right) + N +o(N)
\end{aligned}
\end{equation}
Hence,
\begin{equation}
S[\tilde{\rho}_{\bar E,V,N}] \asymp
\ln\!\left(
\left(\frac{4\pi m e E}{3N}\right)^{3N/2} \cdot\left(\frac{V}{N}\right)^N
\right) + N.
\end{equation}

\subsection{$(E,\bar V,N)$-Ensemble}
Next, consider $(E,\bar V,N)$-Ensemble, whose exact entropy expression takes the form
\begin{equation}
S[\tilde\rho_{_{E,\bar V,N}}^{\mathrm{eq}}]
=
\ln\!\left(
\frac{2\cdot\pi^{3N/2}(2mE)^{(3N-1)/2}}{\Gamma(3N/2)}
\right)
+(N+1)(1+\ln V-\ln (N+1)).
\end{equation}
Since asymptotic limit of the first term has already been evaluated in $(E,V,N)$-Ensemble, it remains to show that
\begin{equation}
(N+1)(1+\ln V-\ln (N+1))\asymp \ln \left(\frac{V}{N}\right)^N+N.
\end{equation}
Obviously,
\begin{equation}
\begin{aligned}
\lim_{E,V,N\to \infty}(N+1)\bigl(1+\ln V-\ln(N+1)\bigr)
&=
(N+1)
\left(
1+\ln\!\left(\frac{V}{N+1}\right)
\right) \\
&=
N\ln\!\left(\frac{V}{N}\right)+N+o(N),
\end{aligned}
\end{equation}
therefore, it immediately follows that
\begin{equation}
S[\tilde{\rho}_{E,\bar V,N}] \asymp N
\ln\!\left(
\left(\frac{4\pi m e E}{3N}\right)^{3N/2} \cdot\left(\frac{V}{N}\right)^N
\right) + N.
\end{equation}

\subsection{$(E,V,\bar N)$-Ensemble}
Finally, we consider the $(E,V,\bar N)$-Ensemble, whose exact entropy does not admit a closed-form expression and is instead given implicitly by
\begin{equation}
S[\tilde{\rho}_{E,V,
\bar N}]=\ln\!\left(
\sum_{n=1}^{\infty}
\frac{2\cdot\pi^{3n/2}(2mE)^{(3n-1)/2}}{\Gamma(3n/2)}\,\frac{V^n}{n!}\,e^{-\beta (n-N)} \right),
\end{equation}
where $N=-\frac{\partial}{\partial \beta}\ln \left(\sum_{n=1}^\infty \frac{2\cdot\pi^{3n/2}(2mE)^{(3n-1)/2}}{\Gamma(3n/2)}\,\frac{V^n}{n!}\,e^{-\beta n}\right)$. Therefore, the asymptotic analysis of $S[\tilde{\rho}_{E,V,
\bar N}]$ is much more complicated than that of the previous ones. For notational convenience, we let
\begin{equation}
X= (2\pi mE)^{3/2}\frac{V}{N^{5/2}}e^{-\beta},
\end{equation}
so that $S[\tilde{\rho}_{E,V,
\bar N}]$ can be rewritten in the form
\begin{equation}
S[\tilde{\rho}_{E,V,
\bar N}]=\ln\!\left(
\frac{2}{\sqrt{2mE}}
\sum_{n=1}^{\infty}
\frac{X^n \cdot N^{5n/2}}
{\Gamma(3n/2)\cdot n!}
\right) + \beta N
\asymp
\ln\!\left(
\sum_{n=1}^{\infty}
\frac{X^n \cdot N^{5n/2}}
{\Gamma(3n/2)\cdot n!}
\right) + \beta N.
\end{equation}
$X$ is determined by the mean particle number constraint:
\begin{equation}
\sum_{n=1}^{\infty}
n\,
\frac{X^n\cdot N^{5n/2}}
{\Gamma(3n/2)\cdot n!}
=
N
\sum_{n=1}^{\infty}
\frac{X^n\cdot N^{5n/2}}
{\Gamma(3n/2)\cdot n!}.
\end{equation}
To show that \(X\) is uniquely determined for any given $N$, define
\begin{equation}
\mathcal A_N(X)
=
\sum_{n=1}^{\infty}
\frac{X^n\cdot N^{5n/2}}
{\Gamma(3n/2)\cdot n!}.
\label{An}
\end{equation}
The mean particle number constraint can be rewritten as
\begin{equation}
\frac{X \mathcal A_N'(X)}{\mathcal A_N(X)}=N.
\label{eq:Xconstraint}
\end{equation}
Introducing
\begin{equation}
M_N(X)
=
\frac{X \mathcal A_N'(X)}{\mathcal A_N(X)},
\end{equation}
we compute
\begin{align}
M_N'(X)
&=
\frac{d}{dX}
\left(
\frac{X \mathcal A_N'(X)}{\mathcal A_N(X)}
\right) \\
&=
\frac1X
\left(
\frac{X^2\mathcal A_N''(X)}{\mathcal A_N(X)}
+
\frac{X\mathcal A_N'(X)}{\mathcal A_N(X)}
-
\left(
\frac{X\mathcal A_N'(X)}{\mathcal A_N(X)}
\right)^2
\right).
\end{align}
Considering $n$ as a random variable and define its probability distribution
\begin{equation}
p_n(X)
=
\frac{X^n\cdot N^{5n/2}}
{\Gamma(3n/2)\cdot n!\,\mathcal A_N(X)},
\end{equation}
we obtain
\begin{equation}
\begin{aligned}
&
\mathcal A'_N(X)=\frac{1}{X}\sum_{n=1}^{\infty}\frac{n\cdot X^n\cdot N^{5n/2}}
{\Gamma(3n/2)\cdot n!}=\frac{\langle n\rangle_X}{X}\mathcal A_N(X),\\
&
\mathcal A''_N(X)=\frac{1}{X^2}\sum_{n=1}^{\infty}\frac{n(n-1)\cdot X^n\cdot N^{5n/2}}
{\Gamma(3n/2)\cdot n!}=\frac{\langle n^2\rangle_X-\langle n\rangle_X}{X^2}\mathcal A_N(X).
\end{aligned}
\end{equation}
Therefore,
\begin{equation}
\begin{aligned}
&
M_N(X)
=
\langle n\rangle_X,\\&
M_N'(X)
=
\frac{\langle n^2\rangle_X-\langle n\rangle_X^2}{X}
=
\frac{\mathrm{Var}_X(n)}{X}.
\end{aligned}
\end{equation}
Since $X= (2\pi mE)^{3/2}\frac{V}{N^{5/2}}e^{-\beta}$ is strictly positive and $\mathrm{Var_X(n)}>0$, we have $M_N'(X)>0$, suggesting that \(M_N(X)\) is strictly increasing for \(X\in(0,\infty)\). By monotonicity, the equation
\begin{equation}
M_N(X)=N
\end{equation}
admits exactly one solution. Later we will show that
\begin{equation}
\lim_{N\to \infty} X=\left(\frac{3}{2}\right)^{3/2}.
\end{equation}
Now, to make progress, we first examine the dominant term in the summation \eqref{An} as \(N\to\infty\). Denote
\begin{equation}
\mathcal B_n
=
\frac{X^n\cdot N^{5n/2}}
{\Gamma(3n/2)\cdot n!}.
\end{equation}
Using Stirling's formula, we obtain
\begin{equation}
\begin{aligned}
\ln \mathcal B_n
&=
n\ln X+\frac{5n}{2}\ln N
-\ln\Gamma\!\left(\frac{3n}{2}\right)
-\ln n! \\
&=
n\ln X+\frac{5n}{2}\ln N
-\frac{3n}{2}\ln\!\left(\frac{3n}{2}\right)
+\frac{3n}{2}
-n\ln n+n
+o(n).
\end{aligned}
\label{lnBn}
\end{equation}
The location of the maximal term is determined by 
\begin{equation}
\begin{aligned}
\frac{d}{dn}\ln \mathcal B_n=0,\quad\frac{d^2}{dn^2}\ln \mathcal B_n<0.
\end{aligned}
\end{equation} 
Differentiating the leading-order expression gives
\begin{equation}
\begin{aligned}
&
\frac{d}{dn}\ln \mathcal B_n
=
\ln X
+\frac{5}{2}\ln N
-\frac{3}{2}\ln\!\left(\frac{3n}{2}\right)
-\ln n
+o(1),\\
&
\frac{d^2}{dn^2}\ln \mathcal B_n =-\frac{5}{2n}<0.
\end{aligned}
\end{equation}
The maximal term is uniquely determined by
\begin{equation}
\ln X
+\frac{5}{2}\ln N
-\frac{3}{2}\ln\!\left(\frac{3n^*}{2}\right)
-\ln n^*
=0,
\end{equation}
which is
\begin{equation}
n^*
=
N
\left(
X\left(\frac{2}{3}\right)^{3/2}
\right)^{2/5}.
\label{n_max}
\end{equation}
Here we introduce the ansatz that $X$ remains finite and strictly positive in the thermodynamic limit, which we will verify later. Under this ansatz, the dominant contribution to the sum arises from a neighborhood of the unique maximizer $n=n^*$. To rigorously show this, we split $\mathcal{A}_N(X)$ into two parts:
\begin{equation}
\sum_{n=1}^{\infty}
\frac{X^n\cdot N^{5n/2}}
{\Gamma(3n/2)\cdot n!}
=
\sum_{n=1}^{[n^*]}
\frac{X^n\cdot N^{5n/2}}
{\Gamma(3n/2)\cdot n!}
+
\sum_{[n^*]+1}^{\infty}
\frac{X^n\cdot N^{5n/2}}
{\Gamma(3n/2)\cdot n!}.
\end{equation}
The first part admits an upper bound:
\begin{equation}
\sum_{n=0}^{[n^*]}\mathcal B_n
\leq
([n^*]+1)\mathcal B_{n*}
\leq
(1+o(1))n^* \mathcal B_{n*}.
\label{first}
\end{equation}
For the second part, introduce the variable $y={n}/{N}$. When $N\to\infty$ and $n
>[n^*]$, one has
\begin{equation}
\mathcal B_n
=
\lim_{N\to \infty} \frac{X^n\cdot N^{5n/2}}{\Gamma(3n/2)\cdot n!}
=
\exp\!\left(
N f_X(y)+o(N)
\right),
\end{equation}
where
\begin{equation}
f_X(y)
=
y\ln X
-\frac52 y\ln y
-\frac32 y\ln\frac32
+\frac52 y .
\end{equation}
In the limit $N\to\infty$,
\begin{equation}
\lim_{N\to \infty}N\sum_{n=[n^*]+1}^{\infty} \frac{1}{N} \mathcal B_n
\asymp
N
\int_{([n^*]+1)/N}^{\infty}
\exp\!\left(
N f_X(y)
\right)dy
<
N
\int_{0}^{\infty}
\exp\!\left(
N f_X(y)
\right)dy
\label{relax}
\end{equation}
The function \(f_X(y)\) has a unique maximum at \(y^*=n^*/N\). Indeed,
\begin{equation}
\begin{aligned}
f_X'(y)
=
\ln X
-\frac52\ln y
-\frac32\ln\frac32,\quad
f_X''(y)
=
-\frac{5}{2y}<0.
\end{aligned}
\end{equation}
To estimate the integral on the right-hand side of \eqref{relax}, we split the integral into two parts:
\begin{equation}
N\int_{0}^{\infty}
e^{Nf_X(y)}\,dy
=
N\int_{0}^{R}
e^{Nf_X(y)}\,dy
+
N\int_{R}^{\infty}
e^{Nf_X(y)}\,dy,
\end{equation}
which holds for arbitrary $R$. We first consider the tail integral. Since
\begin{equation}
\lim_{y\to\infty}f_X(y)
=
-\frac52y\ln y
+O(y),
\end{equation}
there exist constants \(R>\max \{1,y^*\}\) and \(c>0\) such that
\begin{equation}
f_X(y)\le -cy\ln y,
\quad \forall\,
y\ge R.
\end{equation}
For such $R$ and $c$,
\begin{equation}
0
\le
N\int_R^\infty
e^{Nf_X(y)}\,dy
\le
N\int_R^\infty
e^{-Ncy\ln y}\,dy
\le
N\int_R^\infty
e^{-(Nc\ln R)\,y}\,dy
=
\frac{e^{-NcR\ln R}}{c\ln R},
\end{equation}
which becomes $0$ when $N\to \infty$. Therefore, the dominant contribution to the total integral comes from the head integral. To evaluate the head integral, we invoke Laplace's method, which states that, on a finite integration interval containing a unique interior maximizer $y^*$, the leading asymptotic contribution is determined by the value and the local curvature of the exponent at $y^*$\cite{Merhav2010}. Therefore, for the head integral, we obtain
\begin{equation}
N\int_0^R e^{Nf_X(y)}\,dy
\asymp
N e^{Nf_X(y^*)}
\sqrt{\frac{2\pi}{N|f_X''(y^*)|}}
=
e^{Nf_X(y^*)}
\sqrt{\frac{4\pi Ny^*}{5}}.
\end{equation}
Hence
\begin{equation}
N\int_0^\infty
e^{Nf_X(y)}\,dy
\asymp
e^{Nf_X(y^*)}
\sqrt{\frac{4\pi Ny^*}{5}}=O({n^*}^{1/2})\,\mathcal B_{n^*}.
\label{second}
\end{equation}
Combining \eqref{first} and \eqref{second}, we have
\begin{equation}
\begin{aligned}
&
\sum_{n=1}^{\infty}
\frac{X^n\cdot N^{5n/2}}
{\Gamma(3n/2)\cdot n!}
\le
O({n^*}^{1/2})\,\mathcal B_{n^*} + (1+o(1))n^*\mathcal B_{n^*}
=
(1+o(1))n^*\mathcal{B}_{n^*},
\\&
\ln{\left(\sum_{n=1}^{\infty}
\frac{X^n\cdot N^{5n/2}}
{\Gamma(3n/2)\cdot n!}\right)}
\le
\ln n^* + \ln\mathcal{B}_{n^*} + o(1) \asymp \ln\mathcal{B}_{n^*}.
\label{upper_bound}
\end{aligned}
\end{equation}
Thereby, we have rigorously proved that the sum $\mathcal A_N(X)$ is dominated by its maximal term $\mathcal B_{n^*}$, namely the concentration of measure. We now use this result to verify the ansatz introduced above by proving that $\lim_{N\to \infty} X=\left(\frac{3}{2}\right)^{3/2}$. To this end, we reconsider the mean particle number constraint
\begin{equation}
N\cdot\mathcal A_N(X)=\sum_{n=1}^{\infty} n\cdot
\frac{X^n\cdot N^{5n/2}}
{\Gamma(3n/2)\cdot n!}=\sum_{n=1}^{\infty}
\frac{X^n\cdot N^{5n/2}}
{\Gamma(3n/2)\cdot (n-1)!}.
\end{equation}
To identify the dominant term of the right hand side, we denote
\begin{equation}
\mathcal C_n
=
\frac{X^n\cdot N^{5n/2}}
{\Gamma(3n/2)\cdot (n-1)!}.
\end{equation}
Using Stirling's approximation, we obtain
\begin{equation}
\begin{aligned}
\frac{d}{dn}\ln \mathcal C_n
&\asymp
\ln X
+\frac{5}{2}\ln N
-\frac{3}{2}\ln\!\left(\frac{3n}{2}\right)
-\ln (n-1)
\\
&=
\ln X
+\frac{5}{2}\ln N
-\frac{3}{2}\ln\!\left(\frac{3n}{2}\right)
-\ln (n)+\ln(\frac{n}{n-1}).
\end{aligned}
\label{Cn}
\end{equation}
When $N$ is sufficiently large, the maximum point of $\mathcal C_n$, denoted as $n^\star$, will asymptotically coincide with the maximum point of $\mathcal B_n$, i.e.
\begin{equation}
n^\star = (1+o(1))n^*.
\end{equation}
Therefore,
\begin{equation}
N\cdot\mathcal A_N(X)
\asymp
n^\star\cdot \mathcal A_N(X)
\asymp
n^*\cdot \mathcal A_N(X),
\end{equation}
which implies
\begin{equation}
\begin{aligned}
&N=(1+o(1))n^\star=(1+o(1))n^*,\\
&X=(1+o(1))\cdot\left(\frac{3}{2}\right)^{3/2},\\
&\beta=\ln \left( \Big(\frac{4\pi mE}{3N}\Big)^{3/2} \Big(\frac{V}{N}\Big)\right).
\end{aligned}
\end{equation}
Inserting into \eqref{lnBn}, we have:
\begin{equation}
\ln\mathcal A_X(N)=\ln\left( \sum_{n=1}^{\infty}
\frac{X^n\cdot N^{5n/2}}
{\Gamma(3n/2)\cdot n!}\right)
\asymp
\ln\mathcal B_{n^*}
=
\frac{5n^*(1+o(1))}{2} 
\asymp 
\frac{5N}{2}.
\end{equation}
Finally,
\begin{equation}
\begin{aligned}
S[\tilde{\rho}_{E,V,
\bar N}]
&=
\ln\!\left(
\sum_{n=1}^{\infty}
\frac{2\cdot\pi^{3n/2}(2mE)^{(3n-1)/2}}{\Gamma(3n/2)}\,\frac{V^n}{n!}\,e^{-\beta (n-N)} \right)\\
&= \ln \mathcal A_X(N)\,+\beta N\\
&\asymp N\ln\left(\left(\frac{4\pi m E}{3N}\right)^{3/2}\cdot\frac{V}{N}\right)
+
\frac{5N}{2}
\\
&=N\ln\left(\left(\frac{4\pi m E}{3N}\right)^{3/2}\cdot\frac{V}{N}\right)+N\ln e^{3/2}+N\\
&=\ln\!\left(
\left(\frac{4\pi m e E}{3N}\right)^{3N/2} \cdot\left(\frac{V}{N}\right)^N
\right) + N.
\end{aligned}
\end{equation}
Thereby, we have shown that all four representative ensembles reproduce the same leading-order entropy 
\begin{equation}
S\asymp \ln\!\left(
\left( \frac{4\pi m e E}{3N} \right)^{3N/2} \cdot
\left(\frac{V}{N}\right)^N
\right)+N
\end{equation}
in the thermodynamic limit.

\section{Prospect}
\begin{figure}[H]
    \centering
    \includegraphics[width=0.9\linewidth]{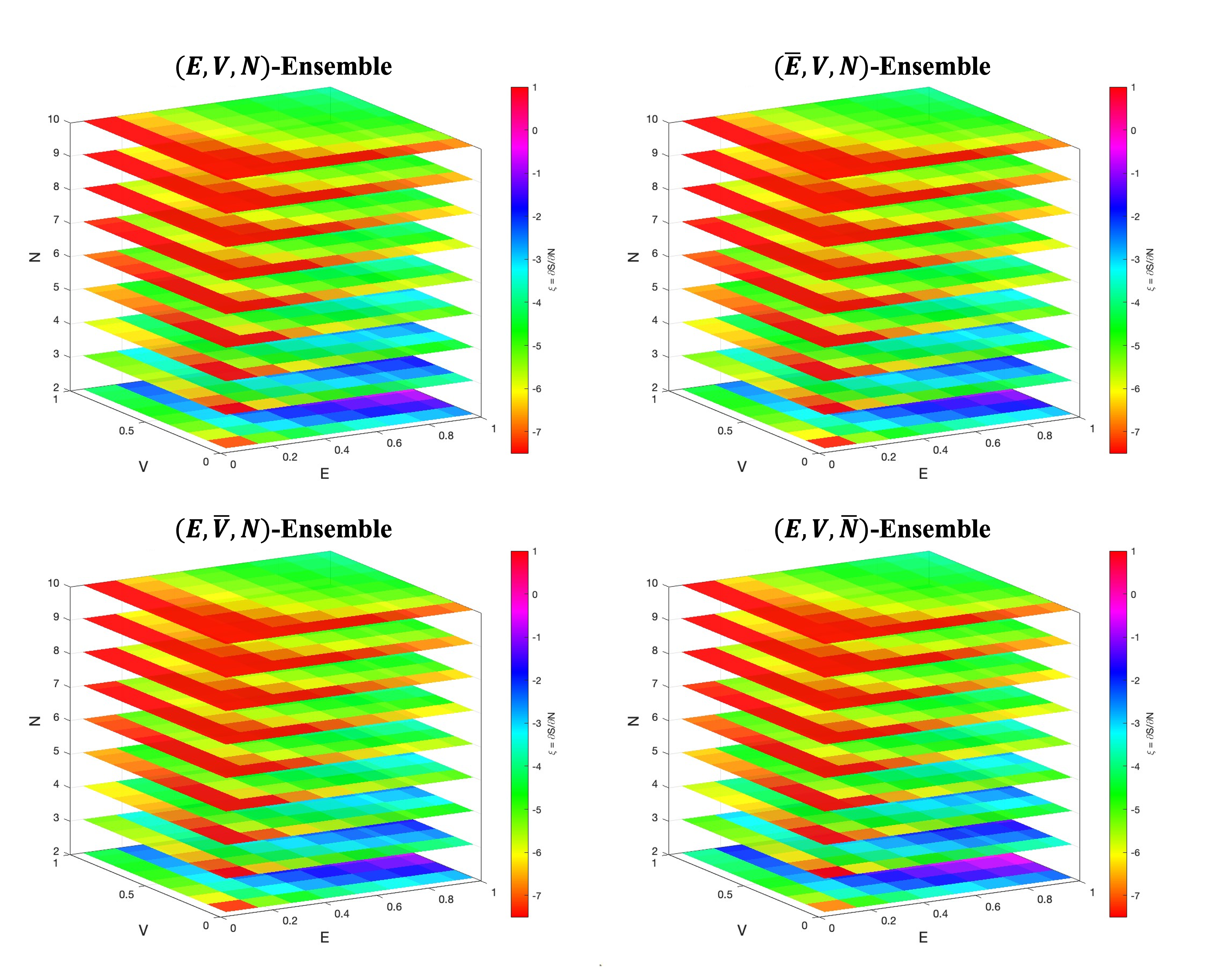}
    \vspace{1em}
    \caption{The values of $\phi$ in the four representative ensembles as functions of $E,V,N$.}
    \label{chem-potential}
\end{figure}

The framework developed in this work provides the foundation for extending equilibrium statistical mechanics beyond the thermodynamic limit. It offers a systematic approach to investigating finite-size thermodynamic effects in nanoscale systems, where the assumptions underlying conventional macroscopic thermodynamics may no longer be valid. One particularly promising application is the thermodynamics of gases confined within nanopores, which can give rise to measurable deviations from bulk behavior. To illustrate the use of the present theory, consider an ideal gas system with energy $E_s$, volume $V_s$, and particle number $N_s$, thermalized in a specific ensemble. Suppose that the system is in contact with a bath, which is also an ideal gas system with energy $E_b$, volume $V_b$, and particle number $N_b$. If the system and bath are allowed to exchange energy, then thermal equilibrium is characterized by the maximization of the total entropy of the combined isolated system. Since the total energy is conserved, the equilibrium condition requires that, for any energy perturbation $\delta E$,
\begin{equation}
S_s(E_s+\delta E,V_s,N_s)
+
S_b(E_b-\delta E,V_b,N_b)
\le
S_s(E_s,V_s,N_s)
+
S_b(E_b,V_b,N_b),
\end{equation}
which implies
\begin{equation}
\begin{aligned}
&
\left(
\frac{\partial S_s}{\partial E_s}
\right)_{V_s,N_s}\cdot d E
-
\left(
\frac{\partial S_b}{\partial E_b}
\right)_{V_b,N_b}\cdot d E=0,
\\
&
\left(
\frac{\partial S_s}{\partial E_s}
\right)_{V_s,N_s}
=
\left(
\frac{\partial S_b}{\partial E_b}
\right)_{V_b,N_b}.
\end{aligned}
\end{equation}
This motivates the definition of
\begin{equation}
\psi
:=
\left(
\frac{\partial S}{\partial E}
\right)_{V,N},
\end{equation}
which behaves as the gauge of thermal equilibrium, determining the direction of spontaneous energy transfer between systems that are not yet in equilibrium. Analogously, we can also define
\begin{equation}
\varphi
:=
\left(
\frac{\partial S}{\partial V}
\right)_{E,N},
\quad
\phi
:=
\left(
\frac{\partial S}{\partial N}
\right)_{E,V},
\end{equation}
to characterize mechanical and chemical equilibrium respectively, while more conventionally, people prefer to use temperature, pressure, and chemical potential as the gauge of thermal, mechanical, and chemical equilibrium:
\begin{equation}
T:=\frac{1}{\psi},
\quad
P:=\frac{\varphi}{\psi},
\quad
\mu:=-\frac{\phi}{\psi}.
\end{equation}
Once the entropy function $S(E,V,N)$ is known exactly, the corresponding temperature, pressure, and chemical potential can be determined exactly through the thermodynamic relations above. Notably, the values of these intensive thermodynamic variables, in fact, depend on the statistical ensemble in which the finite-size system is thermalized. Therefore, determining the appropriate thermalization mechanism, and hence the corresponding statistical ensemble, becomes essential for accurately describing the thermodynamics and kinetics of finite-size systems. For instance, the chemical potential of an ideal gas confined within a nanopore with known energy $E$, volume $V$, and particle number $N$ may deliver quite different values depending on the ensemble in which it is thermalized. More specifically, Fig.~\ref{chem-potential} illustrates this ensemble dependence by comparing
\[
\phi=\left(\frac{\partial S}{\partial N}\right)_{E,V},
\]
which is directly related to the chemical potential, among the four representative ensembles as functions of $E$, $V$, and $N$.

\appendix    
\subsection* {Acknowledgments}
The author is grateful to Professor Vivishek Sudhir at the Massachusetts Institute of Technology for offering the exceptional course \textit{2.S982: Quantum Machines in a Classical World}, which sparked the author's interest in closed dynamical systems. The author sincerely thanks Professor Adam Willard and Professor Iwnetim Abate at the Massachusetts Institute of Technology for their support of this work, as well as Professor Amr Dodin at The Ohio State University for his valuable feedback. The author also thanks the members of the Willard and Abate groups for many helpful discussions.


\bibliography{report}   
\bibliographystyle{spiejour}   


\end{spacing}
\end{document}